\documentclass[%
showpacs,
 amsmath,amssymb,
 aps,
pra,
]{revtex4-1}

\usepackage{graphicx}
\usepackage{bm}

\begin{document}


\title{Trapping and reshaping of low-intensity radiations by soliton trains in gas-filled hollow-core photonic crystal fibers.}


\author{R. D. Dikand\'e Bitha}
\email{rodrigue.donald@ubuea.cm}
\affiliation{Laboratory of Research on Advanced Materials and Nonlinear Sciences (LaRAMaNS), Department of Physics, Faculty of Science, University of Buea P.O. Box 63 Buea, Cameroon}

\author{D. S. Mbieda Petmegni}
\email{mbiedaduplex@gmail.com}
\affiliation{Laboratory of Research on Advanced Materials and Nonlinear Sciences (LaRAMaNS), Department of Physics, Faculty of Science, University of Buea P.O. Box 63 Buea, Cameroon}

\author{Alain M. Dikand\'e}
\email{dikande.alain@ubuea.cm}
\affiliation{Laboratory of Research on Advanced Materials and Nonlinear Sciences (LaRAMaNS), Department of Physics, Faculty of Science, University of Buea P.O. Box 63 Buea, Cameroon}

\date{\today}

\begin{abstract}
An optical trapping scheme is proposed by which ultrashort low-amplitude radiations, co-propagating with a continuous train of temporal pulses in a hollow-core photonic crystal fiber filled with Raman-inactive noble gases, can be trapped and reshaped into optical soliton trains by means of cross-phase modulation interactions. The scheme complements and extends a recently proposed idea that a single-pulse soliton could trap an ultrashort small-amplitude
radiation in a symmetric hollow-core photonic crystal fiber filled with a noble gas, thus preventing its dispersion [M. F. Saleh and F. Biancalana, Phys. Rev. A87, 043807 (2013)]. We find a family of three distinct soliton-train boundstates with different propagation constants, one being a "duplicate" of the trapping pulse train. We analyze the effects of self-steepening on the trapping (i.e. pump) and trapped (i.e. probe) field profiles and find that self-steepening causes a uniform shift in position of the pump soliton train, but a complex motion for the probe dominanted by anharmonic oscillations of their temporal positions and phases. The new trapping scheme is intended for optical applications involving optical-field cloning and duplication via wave-guided-wave processes, in photonic fiber media in which interplay time-division multiplexed high-intensity pulses coexisting with continuous-wave radiations.
\end{abstract}


\maketitle

\section{\label{sec:one} Introduction}
Solitons in optical media are formed when the group-velocity dispersion (GVD) is compensated by the self-phase modulation (SPM) favored by an intensity-dependent refractive index \cite{t1,t2,t3,t4,t4a}. The robustness of the resulting optical field not only ensures propagation over long distance with a permanent shape profile, but can also enable the resulting soliton field to trap and convey radiations of relatively low amplitudes that would not survive in the system because of dispersion. Most commonly optical trapping has been associated with the interplay of a pair of high-intensity soliton-type pulses \cite{m1,m2,m3}, for instance when two solitons propagate in opposite dispersion regimes, the group velocity matching condition is satisfied such that the pair can form a self-trapped bound state. This is the case for example with femtosecond red-shifted solitons due to intra-pulse Raman scatterings, which has been shown to trap wavelength weak blue-shifted nonlinear dispersion wave-packets across the zero dispersion regime \cite{t5,t6,t7,t8,t9}. Optical trapping can also occur through the nonlinear cross-phase modulation (XPM) effect between different pulses, when group velocities of two optical solitons are matched \cite{t10}. There has equally been interest in situations where a bright soliton is trapped in the normal dispersion regime, when coupled to a dark soliton in the anomalous dispersion regime through the nonlinear XPM effect \cite{t13,t14,t15}. In this work we focus on a distinct optical trapping scheme involving a strong pump field and small-amplitude radiations \cite{steig1,steig2,steig3,t12}, both propagating along a photonic crystal fiber (PCF). \par
The advent of PCFs has been a significant advance in the field of nonlinear optics. Indeed they offer many degrees of freedom in their design which allows achieve a variety of peculiar properties, making them interesting for a wide range of applications \cite{t16,t17,t18}. Hollow-Core PCFs (HCPCFs) in particular \cite{t14,moha,fac,moha1}, that guide light essentially within a hollow region such that only a minor portion of the optical power propagates in the solid fiber material, have been developed to probe the nonlinear interaction of light in a new wide range of distinct media. Among several singular features it has been shown that they can support periodic soliton trains \cite{t12,t19}, and when filled with a Raman-active gas it is possible to trap a periodic train of optical pulses forming a soliton crystal \cite{t19}. \par
HCPCFs, filled with Raman active gases to reduce the Raman threshold, have recently been successfully used to investigate light-matter interactions in the presence of different noble gases \cite{t20}. By applying intense broadband pulses to access the ionization regime of the noble gases, blue-shifted self-frequency intrapulses have been observed and thoroughly studied \cite{t21, t22}. Launching of intense narrow band pulses has been shown to lead to ionization-induced asymmetric SPM, and a universal modulational instability \cite{t22}. Most recently \cite{t21}, a HCPCF filled with Raman-inactive noble gas and having a symmetric core was proposed as a perfect host to achieve the trapping of low-intensity radiations by high-intensity pulses both propagating at the same frequency. The two optical fields were assumed to be in two distinct circular polarization states in the deep anomalous dispersion regime of the fiber, but having the same frequency allowed them being in the same group velocity and hence to be naturally confined in the same dispersion regime. Instructively, using circular polarization states to propagate an optical pulse together with a small-amplitude field, offers the relevant advantage of avoiding unwanted additional nonlinear phenomena that would emerge if linearly polarized states were injected in the HCPCF. Thus by assuming an input single-pulse soliton in the regime of symmetric HCPCF, the authors established that any small-amplitude radiation propagating with the pump pulse would be trapped and reshaped into either an even mode with the same shape and GVD as the pump soliton but with a smaller propagation constant, or an odd mode with the same GVD as the pump but a different shape profile and higher propagation constant \cite{t21}. The even and odd modes found in this previous study were actually two localized modes representing the complete set of boundstates of the eigenvalue problem associated with the scattering of the pump pulse with small-amplitude oscillations, in the symmetric-core HCPCF filled with inactive-Raman noble gases. In the same context it was found \cite{t21} that the account of self-steepening \cite{mar,stepe1,stepe2} leads to an osccillating motion of the trapped optical field, similar to the motion of a mechanical pendulum.  \par
Albeit the trapping scheme proposed in ref. \cite{t21} is relevant for physical contexts where the pump field is a single-pulse soliton, for HCPCFs specifically designed to support periodic structures such as pulse trains or multiple-soliton complexes in general, the scheme turns out to be limited. In this work we extend the idea of ref. \cite{t21} to this later context, by considering a train of time entangled pulses forming a periodic soliton crystal propagating together with small-amplitude radiations in the anomalous dispersion regime of a HCPCF. \par 
In sec. \ref{sec:two} we construct a multi-soliton complex by defining and ansatz representing time-division multiplexed temporal pulse solitons, and shows that such artificial structure forming a periodic soliton crystal is solution to the NLSE. Then by considering this multi-soliton complex as the pump field, we formulate the scattering problem corresponding to its interactions with small-amplitude radiations. We derive the associated boundstate spectrum which consists of exactly three self-trapped soliton-train structures, one of which is a duplicate of the pump field while two are nearly degenerates. In sec. \ref{sec:three} we analyze the effect of self-steepening, related to optical shocks, on characteristic parameters of the pump pulse train as well as of three self-trapped modes. By means of a variational approach based on a Lagrangian formalism for characteristic positions and phases of the self-trapped modes, we derive three sets of two coupled first-order ordinary differential equations for variational parameters. These equations are solved numerically following a six-order Runge-Kutta scheme adapted from Luther \cite{luth}. We end with sec. \ref{sec:four} where concluding remarks are presented.

\section{\label{sec:two} Model and self-trapped periodic soliton trains}
 Consider a periodic train of intense pulses propagating in the background of low-amplitude light radiations in a HCPCF filled with a noble gas. We assume the pulse train and radiations have the same central frequency but different circular polarization states, and that the radiations are of very small amplitudes compared with the amplitude of the pump pulse train. Since we postulate that the two fields propagate coupled one to the other through a cross-phase modulation nonlinearity \cite{steig1,steig2,steig3,t12}, the propagation equations for the two fields can be represented by two coupled nonlinear Schr\"odinger equations of the forms \cite{t21}:

\begin{eqnarray}
i\partial_z B_1  + \hat{D}_1 (i\partial_t )B_1  + \frac{2}{3}\gamma \vert B_1 \vert^2 B_1  &=& \frac{i}{2}\Delta \beta B_2, \label{1} \\
i\partial_z B_2  + \hat{D}_2 (i\partial_t )B_2  + \frac{4}{3}\gamma \vert B_1 \vert^2 B_2  &=& \frac{i}{2}\Delta \beta B_1, \label{2}
\end{eqnarray}

Where 
\begin{equation}
\hat{D}_j (i\partial_t ) \equiv i\beta_{1j}\partial_t - \frac{1}{2}\beta_{2j}\partial_t^2. \label{1a}
\end{equation}

$z$ and $t$ in eqs. (\ref{1}) and (\ref{2}) refer to the longitudinal coordinate along the fiber and the local time respectively, $B_{1,2}(z,t)$ are the slowly varying amplitudes of the fields (with $B_{1}$ corresponding to the pump field and $B_{2}$ to the small-amplitude radiation), $\beta_{1j}$ and $\beta_{2j}$ are their first and second-order dispersion coefficients $(j = 1,2)$ and $\Delta \beta$ is the fiber birefringence. We assume the medium is Raman and plasma free, and consider a PCF with a perfectly symmetric core such that $\beta_{11} = \beta_{12}$, $\beta_{21} = \beta_{22}$ and $\Delta\beta = 0$. The third term in eq. (\ref{1}) accounts for the SPM effect while the third term in eq. (\ref{2}) characterizes the XPM interaction between the fields. We introduce new (dimensionless) variables i.e. $\xi = z/z_{0}$, $\tau = t/t_{0}$, $U_{1} = B_{1}/B_{0}$ and $U_{2} = B_{2}/B_{0}$ where $B_{0}^{2} = 3/(2\gamma z_{0})$, $z_{0} = t_{0}^{2}/\left| {\beta_{21} } \right|$ and $t_{0}$ is a reference time. Defining the group velocity as $v_{g} = 1/\beta_{11}$, the coupled set eqs. (\ref{1})-(\ref{2}) can be transformed into:
\begin{eqnarray}
i\partial_\xi U_1  + \frac{1}{2}\partial_{\tau \tau} U_1  + \vert  U_1 \vert^2 U_1  &=& 0, \label{3} \\
i\partial_\xi U_2 + \frac{1}{2}\partial_{\tau \tau} U_2  + \vert  U_1 \vert^2 U_2  &=& 0. \label{4}
\end{eqnarray} 
Eq. (\ref{3}) is the cubic nonlinear Schr\"odinger equation (NLSE), its soliton solution describing a localized pulse was obtained in \cite{t21} as beeing: 
\begin{equation}
B_1(\xi, \tau)=\eta\,sech\left(\eta\tau \right)\exp{(i\eta^2\xi/2)}. \label{sing}
\end{equation}
In concrete terms this solution represents a single-pulse temporal soliton undergoing a spatial modulation upon propagation along the HCPCF, with a modulation constant $q=\eta^2/2$. The quantity $\eta$ is the amplitude of the temporal soliton, as it is apparent $\eta$ also fixes the width at half height of the pulse. \\
From a mathematical standpoint the existence of a single-pulse solution such as (\ref{sing}) involves boundary conditions for which the wave wings are required to vanish asymptotically as $\tau\rightarrow \infty$, while the pulse shape sharpens at $\tau=0$. If we instead require the pulse profile to fully deploy over a finite time interval, we can envisage the possibility to accommodate several pulses over a finite time interval. In this later case we can envisage a soliton train structure consisting of time-entangled pulses similar to (\ref{sing}), forming a periodic temporal soliton crystal whose amplitude can formally be expressed \cite{men1}:
\begin{equation}
U_1(\xi,\tau)= \sum_{n=0}^{M}{b_n sech\left[\sigma_n\left(\tau - n\,\tau_0\right)\right]}, \label{temp}
\end{equation}
in which $b_n$ is the amplitude and $\sigma_n$ is the inverse width of the $n^{th}$ pulse component, and $\tau_0$ is the separation between pulses in the soliton train. In the particular case when the constituent pulses are identical in shapes, amplitudes and widths, the discrete sum in (\ref{temp}) is exact for $M\rightarrow \infty$ \cite{malomed} and leads to the periodic train of localized pulses \cite{t15,t12,t19,t23,t31}:
\begin{equation}
U_1(\xi ,\tau ) = \frac{\eta}{\sqrt {2 - \kappa^2 }}dn\left[ {\frac{\eta}{\sqrt{2 - \kappa^2 }}\tau} \right]e^{i\eta ^2 \xi /2}. \label{5}
\end{equation}
This periodic soliton train is also solution to the NLSE (\ref{3}), with $dn$ here denoting one of four Jacobi Elliptic functions \cite{t26,t27,t28} of modulus $\kappa$ ($0\leq \kappa\leq 1$). $dn$ is a periodic function of its argument $\tau$ with a period: 
\begin{equation}
T_{dn}  = 2K(k)/\sigma, \hskip 0.25truecm \sigma = \frac{\eta}{\sqrt{2 - k^2}}, \label{6}
\end{equation}
where $K(k)$ is the Elliptic integral of the first kind. When $\kappa=1$, $T_{dn}$ tends to infinity and the $dn$ function tends to $sech$ (i.e. the case considered in ref. \cite{t21}). Conversely when $\kappa$ decreases to zero, the period gets smaller and the pulse train decays into a periodic train of small-amplitude (i.e. harmonic) waves. \par
\begin{figure} 
\includegraphics{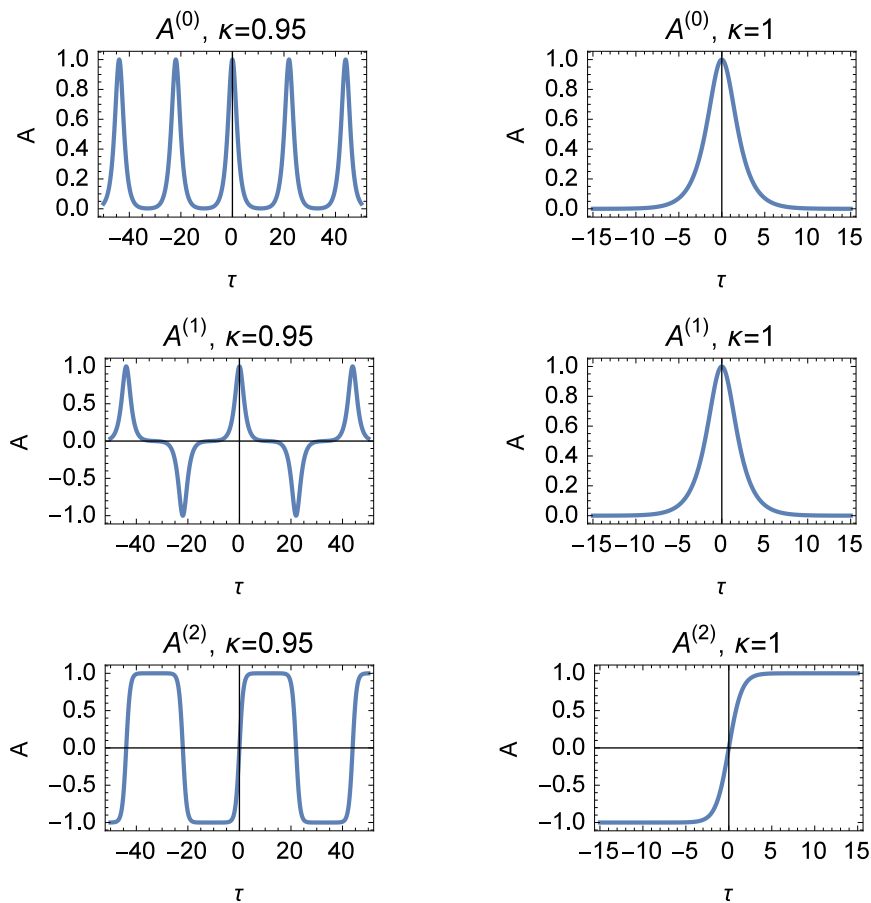}
\caption{\label{fig:one} Temporal profiles of the three soliton-train boundstates (\ref{lam1}), (\ref{lam2}) and (\ref{lam3}). Left graphs are for $\kappa=0.95$ (soliton-train regime), right graphs are for $\kappa=1$ (single-soliton regime).} 
 \end{figure}

Turning to profiles of the radiations propagating bound to the periodic pulse train eq. (\ref{5}), we start by writing the solution to eq. (\ref{4}) in the general form:
\begin{equation}
U_2(\xi ,\tau ) = A(\tau)e^{i\rho\xi}, \label{7}
\end{equation}
describing a radiation field of temporal envelope $A(\tau)$ and propagation constant $\rho$. Substituting this into eq. (\ref{4}) we obtain:
\begin{equation}
\frac{\partial^2 A}{\partial \mu^2} + \left[{2 - \frac{(2 - k^2)}{\eta^2}\rho  - 2\kappa^2\,sn^2(\mu)} \right]A = 0,  \label{8}
\end{equation}
With
\begin{equation}
\mu = \sigma\tau. 
\end{equation}
Eq. (\ref{8}) is a first-order Lam\'e-type \cite{t28} eigenvalue-value problem, whose boundstate spectrum comprises exactly three modes \cite{t28,t29} i.e.:
\begin{eqnarray}
A^{(0)}(\mu) &=& a_0 dn(\mu),\hskip 0.3truecm \rho^{(0)} (\kappa) = \eta^2, \label{lam1} \\
A^{(1)}(\mu) &=& a_1 cn(\mu),\hskip 0.3truecm \rho^{(1)}(\kappa) = \frac{\eta^2}{2-\kappa^2}, \label{lam2} \\
A^{(2)}(\mu) &=& a_2 sn(\mu),\hskip 0.3truecm \rho^{(2)}(\kappa) = \frac{1-\kappa^2}{2-\kappa^2}\eta^2. \label{lam3}
\end{eqnarray}
$a_0$, $a_1$ and $a_2$ in the three last formula are constant amplitudes obtained by normalizing the fields \cite{t29}. Temporal profiles of the three soliton-train boundstates are represented in fig. \ref{fig:one}, for $\kappa=0.95$ (left graphs) and $\kappa=1$ (right graphs). 
It is quite remarkable that the first boundstate is identical in shape with the pump pulse (\ref{5}), suggesting that one of the possible modes generated by self-trapping of small-amplitude radiations in the HCPCF is a "duplicate" of the pulse train. The duplicate moves together with the pump pulse train but with a propagation constant which is twice the one of the pump. The second and third modes are topologically distinct from the first mode, however as $\kappa\rightarrow 1$ the soliton-train boundstates (\ref{lam1}) and (\ref{lam2}) merge into the same boundstate with a "sech" profile and the same propagation constant $\eta^2$. In this limit the third mode reduces to a single temporal dark soliton with a zero propagation constant. 

\section{\label{sec:three} Optical shock: Variational approach}

The account of higher-order nonlinearities and/or nonlinear dispersion terms in optical media, can lead to an optical shock causing a distorsion of pulses as they propoagate in the system. Such phenomenon, called self-steepening \cite{mar,stepe1,stepe2}, has been observed in hollow-core fibers \cite{bej,chad} where they are shown to accelerate pulses (see e.g. \cite{rai}). In the presence of self-steepening, the propagation equations (\ref{3})-(\ref{4}) for the pump and radiation envelopes become:
\begin{eqnarray}
i\partial_\xi U_1 + \frac{1}{2}\partial_{\tau \tau}U_1+\vert U_1 \vert^2 U_1  &=&  - i\tau _{sh} \partial_\tau(\vert U_1 \vert^2 U_1), \label{e1} \\
i\partial_\xi U_2 + \frac{1}{2}\partial_{\tau \tau} U_2  + 2\vert U_1 \vert^2 U_2 &=&  - 2i\tau_{sh} \partial_\tau (\vert U_1 \vert^2 U_2 ), \label{e2}
\end{eqnarray}
where $\tau_{sh} = (\omega_0 t_0)^{-1}$ is the normalized shock coefficient, with $\omega_0$ the central frequency of individual pulses in the time-division multiplexed pulse structure \cite{men1}. We seek for solutions to eqs. (\ref{e1}) and (\ref{e2}) with same profiles as in the absence of self-steepening. For eq.(\ref{e1}) we find that the pulse-train soliton profile obtained in (\ref{5}) is a solution but with a new form: 
\begin{equation}
U_1(\xi ,\tau) = \sigma\,dn\left[\sigma\left(\tau  - \nu \xi\right) \right]e^{i\eta^2\xi/2}, \label{e3a}
\end{equation}
provided $\nu =\tau _{sh}\eta^{2}$. So to say, in the presence of self-steepening the periodic pulse train solution (\ref{5}) undergoes a uniform shift $\tau_{\nu\xi}$ in order to preserve its shape profile. As for the soliton-train boundstates, the effect of self-steepening on their characteristic parameters is not as simple and straightforward as in the case of the pump pulse train as one can check from the corresponding equation (\ref{e2}). Nevertheless, if we assume the optical shock caused by self-steepening to be weak enough such that profiles of the three modes are always preserved, it is ready to look at the self-steepening as a perturbation and hence a reasonably acceptable insight onto their dynamics can be gained within the framework of a variational treatment. \par 
Proceeding with, let us start by expressing the field $U_2(\xi, \tau)$ as:
\begin{equation}
U_2(\xi,\tau) = U_0 A_{1,2,3}\left[\sigma(\tau - \tau_p(\xi)) \right] e^{-i \delta(\xi) \left[\tau - \tau_p(\xi)\right]}, \label{e3b}
\end{equation}
with $A_i\left[x \right] \equiv \{dn(x), sn(x), cn(x)\}$, where $\tau_p$ and $\delta$ are the variational paramaters related to shifts in the temporal and spectral positions of the boundstates. As our objective is to determine the set of differential equations governig the evolution of the two variational parameters, it is useful to remark that the coupled set eqs.~(\ref{e1})-(\ref{e2}) is derived from application of the Euler-Lagrange formalism on the Lagrangian density:
 
\begin{equation}
l_d=-\textit{Im} \lbrack U_2 \partial_{\xi} U_2^* + 2 \varepsilon U_2^* \rbrack + \frac{1}{2} \vert \partial_{\tau} U_1\vert^2
+ 2 \vert U_1 \vert^2 \vert U_2\vert^2, \label{e4}
\end{equation}
where \textit{Im} means the imaginary part of the argument. The corresponding Lagrangian $L$ follows from the formula:
\begin{eqnarray}
L=\int^{{\bf K}}_{{-\bf K}}{l_d\quad d\tau},
\label{e5}
\end{eqnarray}
where we defined ${\bf K}=K(k)$ and the integral is restricted over one period of the soliton train \cite{t31}.

\begin{figure*}\centering
\begin{minipage}[c]{0.5\textwidth}
\includegraphics[width=3.in,height=2.in]{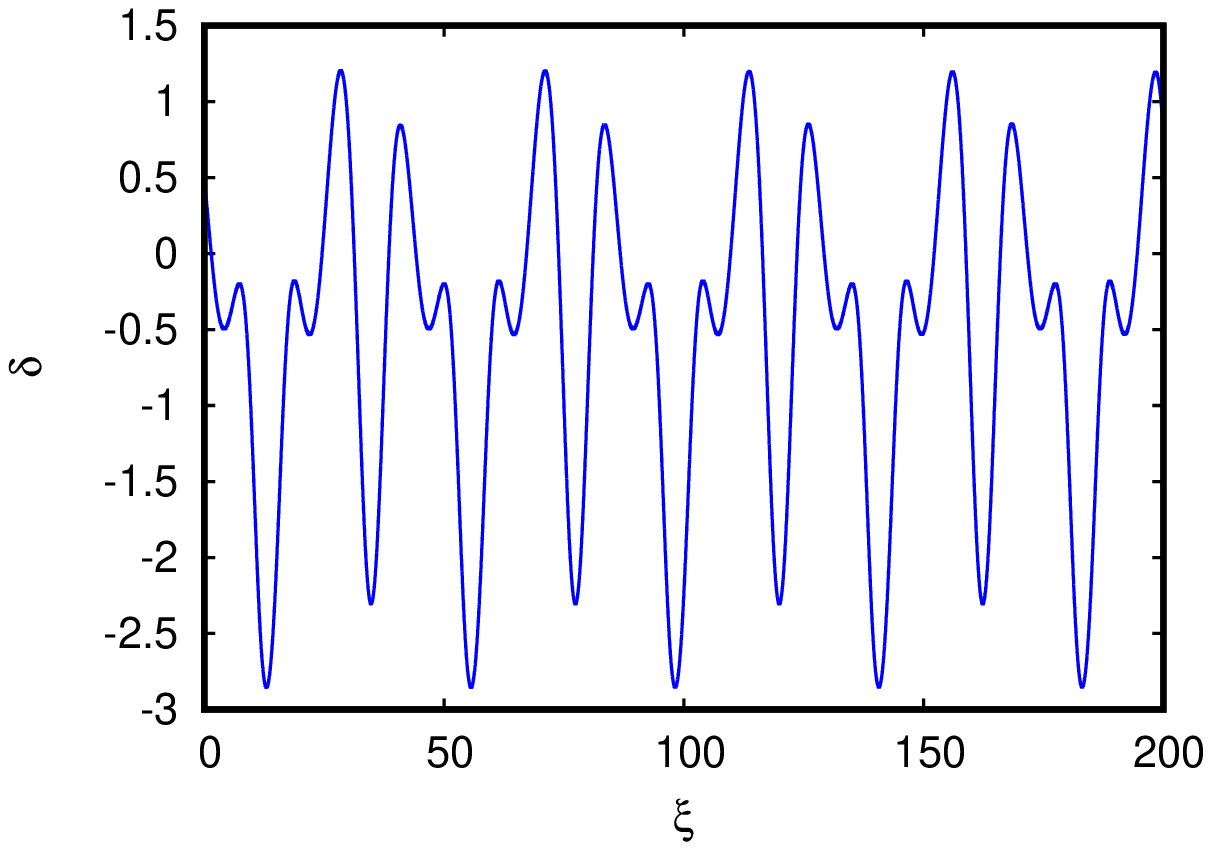}
\end{minipage}%
\begin{minipage}[c]{0.5\textwidth}
\includegraphics[width=3.in,height=2.in]{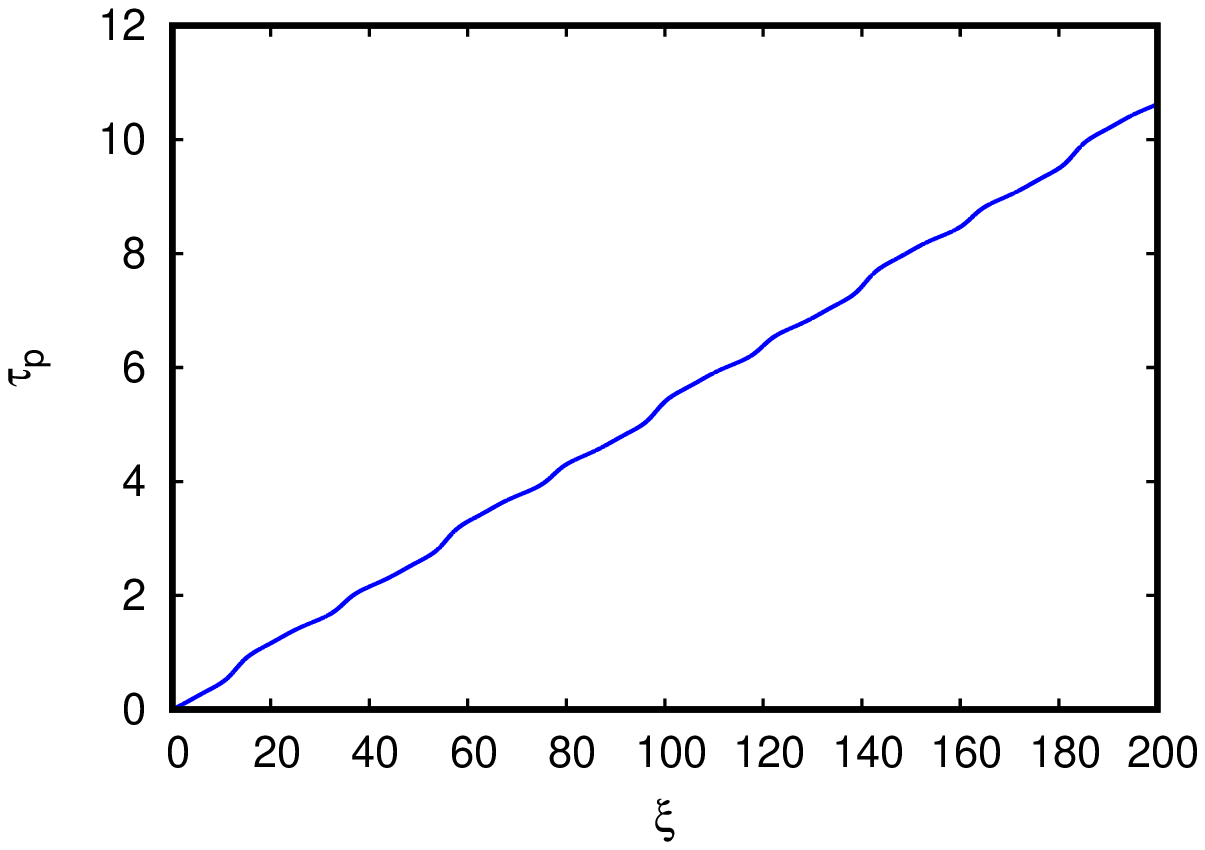}
\end{minipage}\\
\begin{minipage}[c]{0.5\textwidth}
\includegraphics[width=3.in,height=2.in]{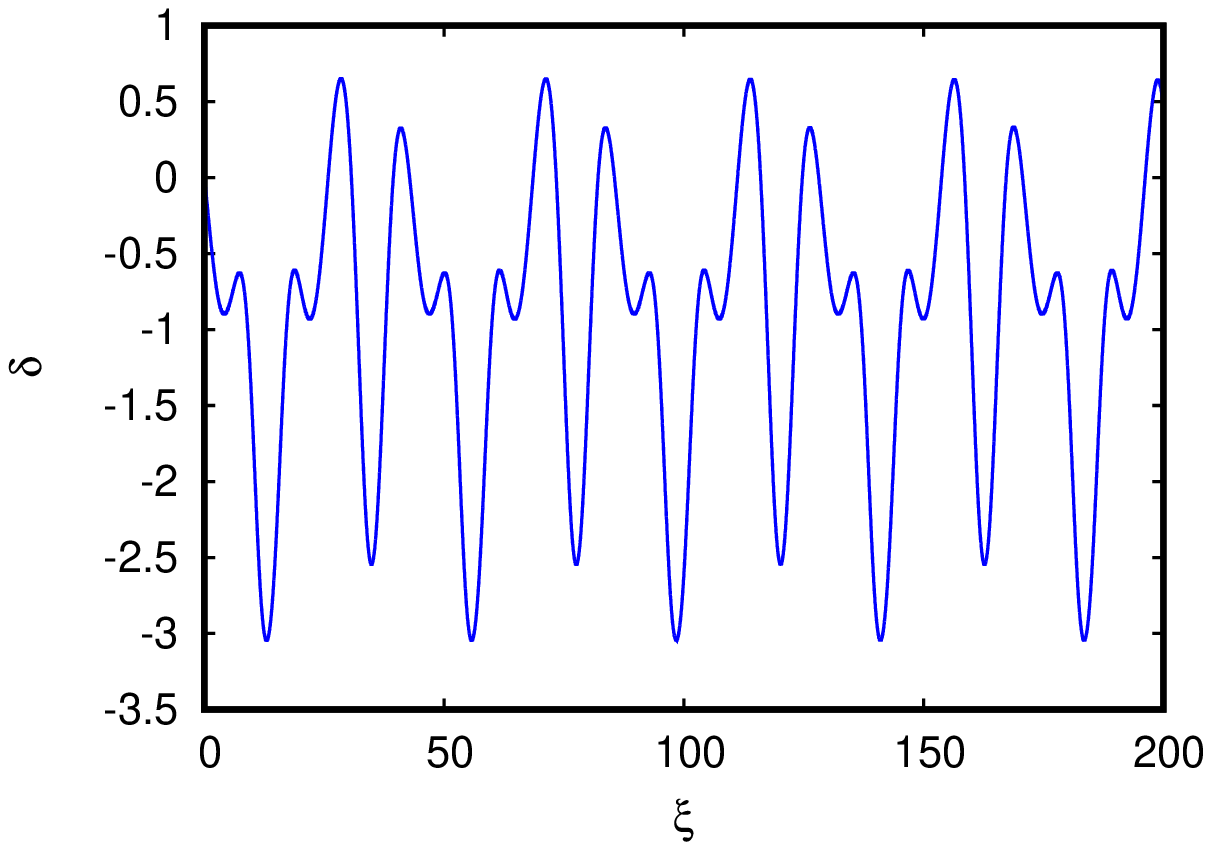}
\end{minipage}%
\begin{minipage}[c]{0.5\textwidth}
\includegraphics[width=3.in,height=2.in]{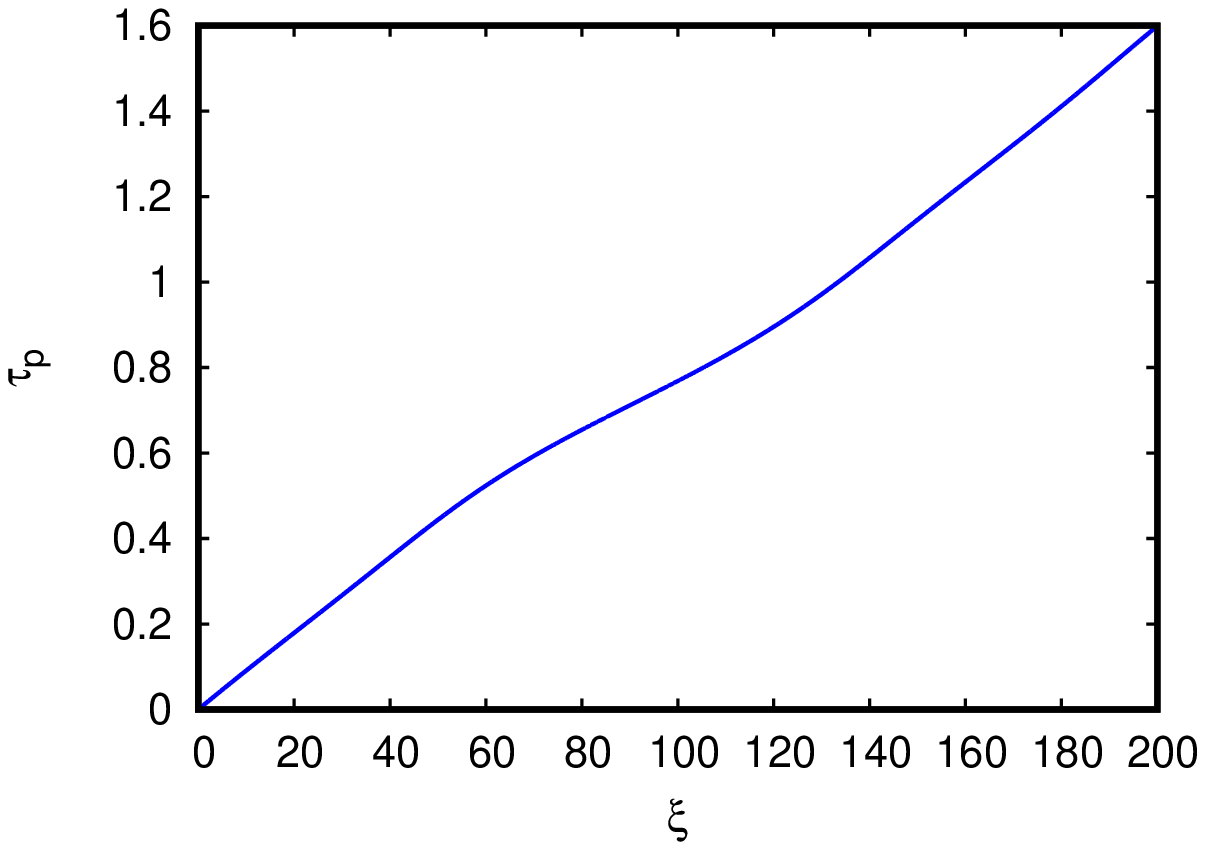}
\end{minipage}\\
\begin{minipage}[c]{0.5\textwidth}
\includegraphics[width=3.in,height=2.in]{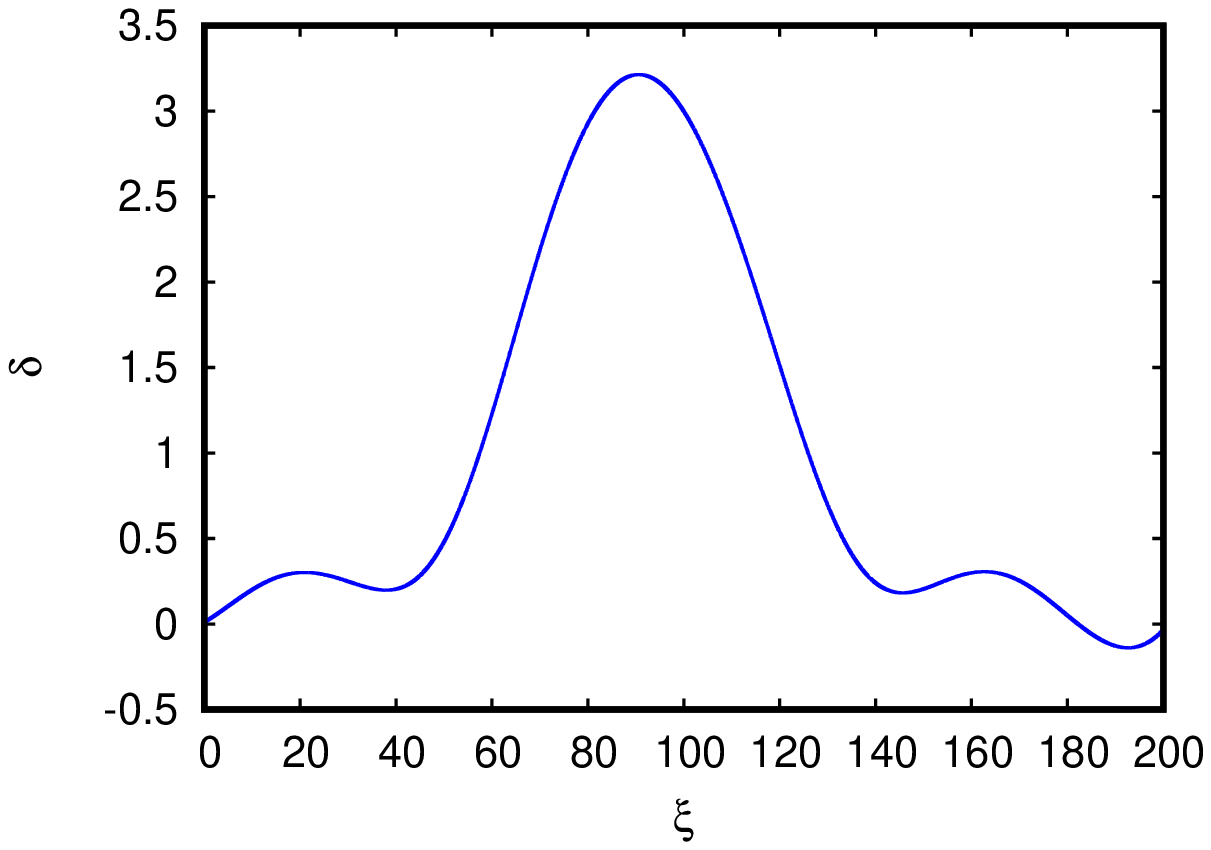}
\end{minipage}%
\begin{minipage}[c]{0.5\textwidth}
\includegraphics[width=3.in,height=2.in]{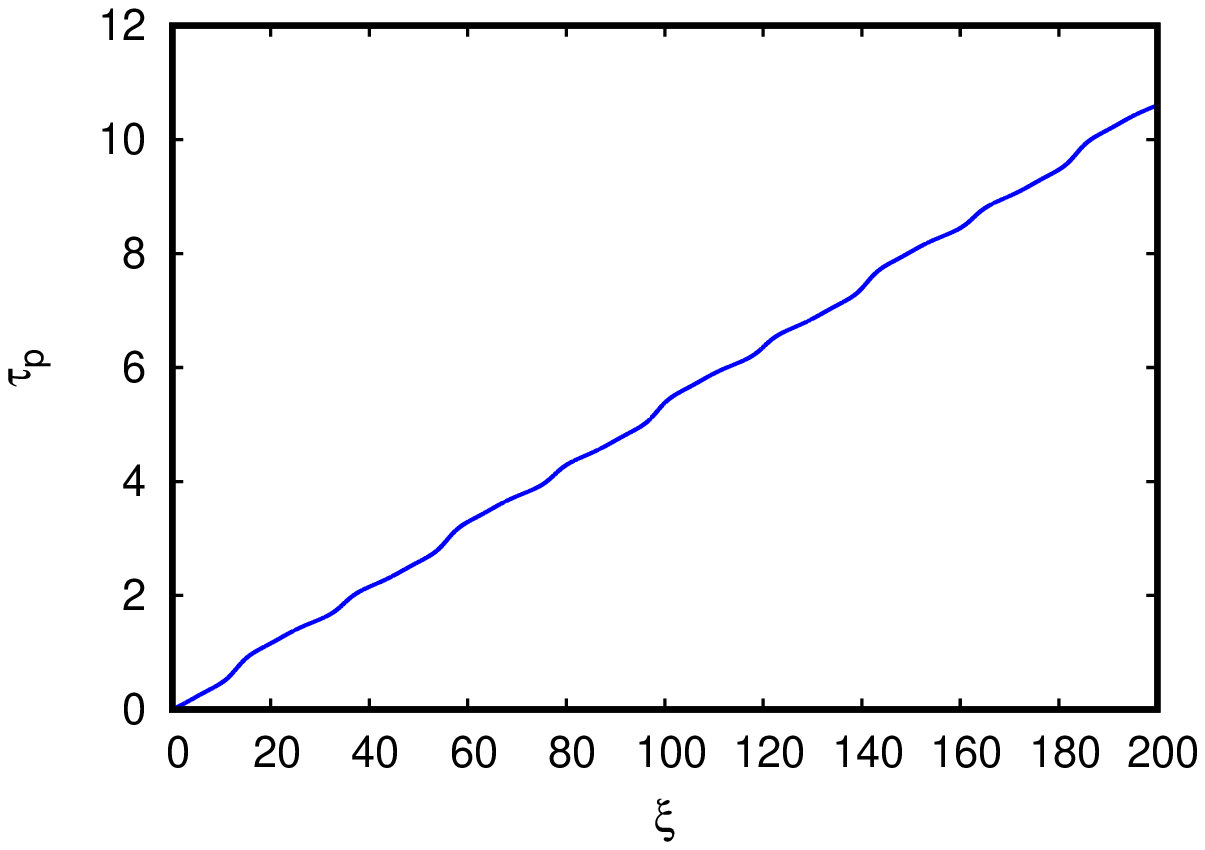}
\end{minipage}
\caption{\label{fig:two} Plots of $\delta$ (right graphs) and $\tau_p$ (left graphs) versus $\xi$, for $\kappa=0.97$, $\eta=0.75$ and $\tau_{sh}=0.08$. Top graphs: $dn$ mode, middle graphs: $cn$ mode, bottom graphs: $sn$ mode.}
\end{figure*}

\begin{figure*}\centering
\begin{minipage}[c]{0.5\textwidth}
\includegraphics[width=3.in,height=2.in]{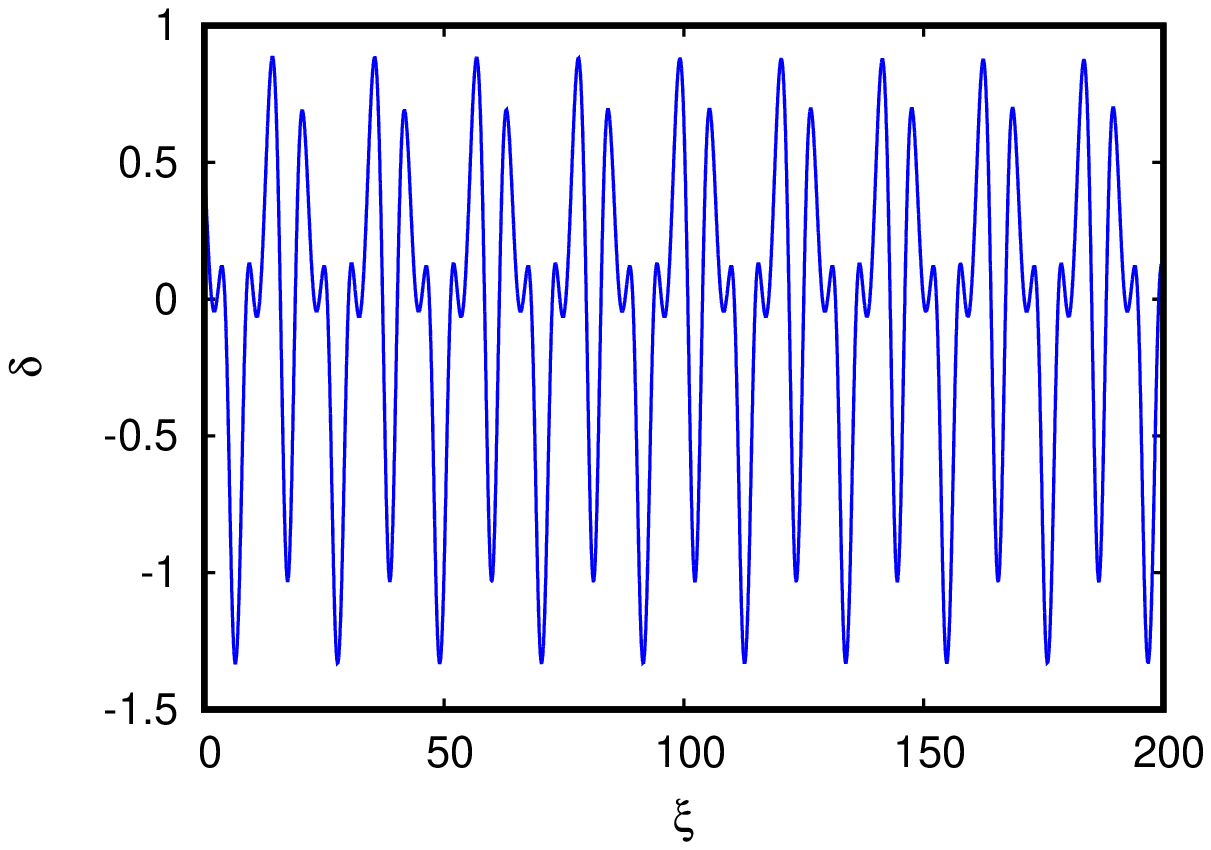}
\end{minipage}%
\begin{minipage}[c]{0.5\textwidth}
\includegraphics[width=3.in,height=2.in]{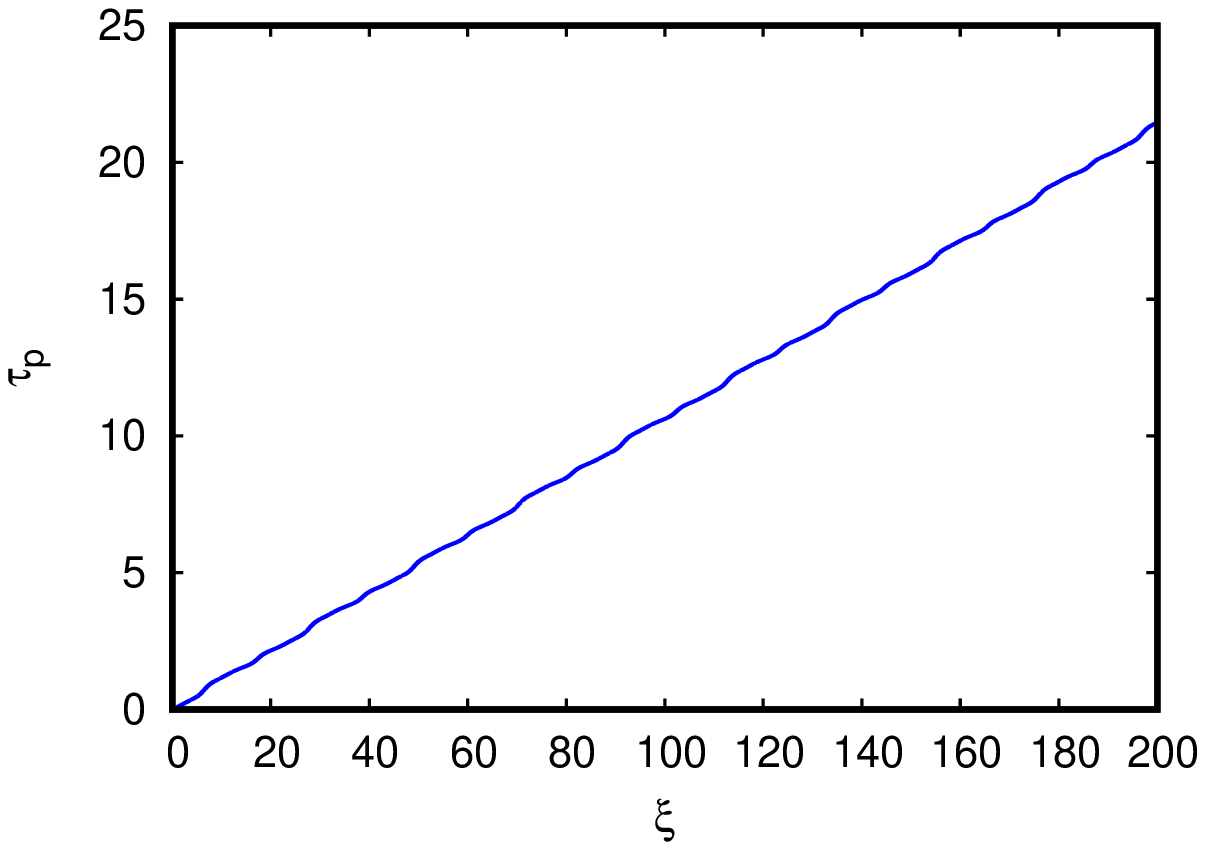}
\end{minipage}\\
\begin{minipage}[c]{0.5\textwidth}
\includegraphics[width=3.in,height=2.in]{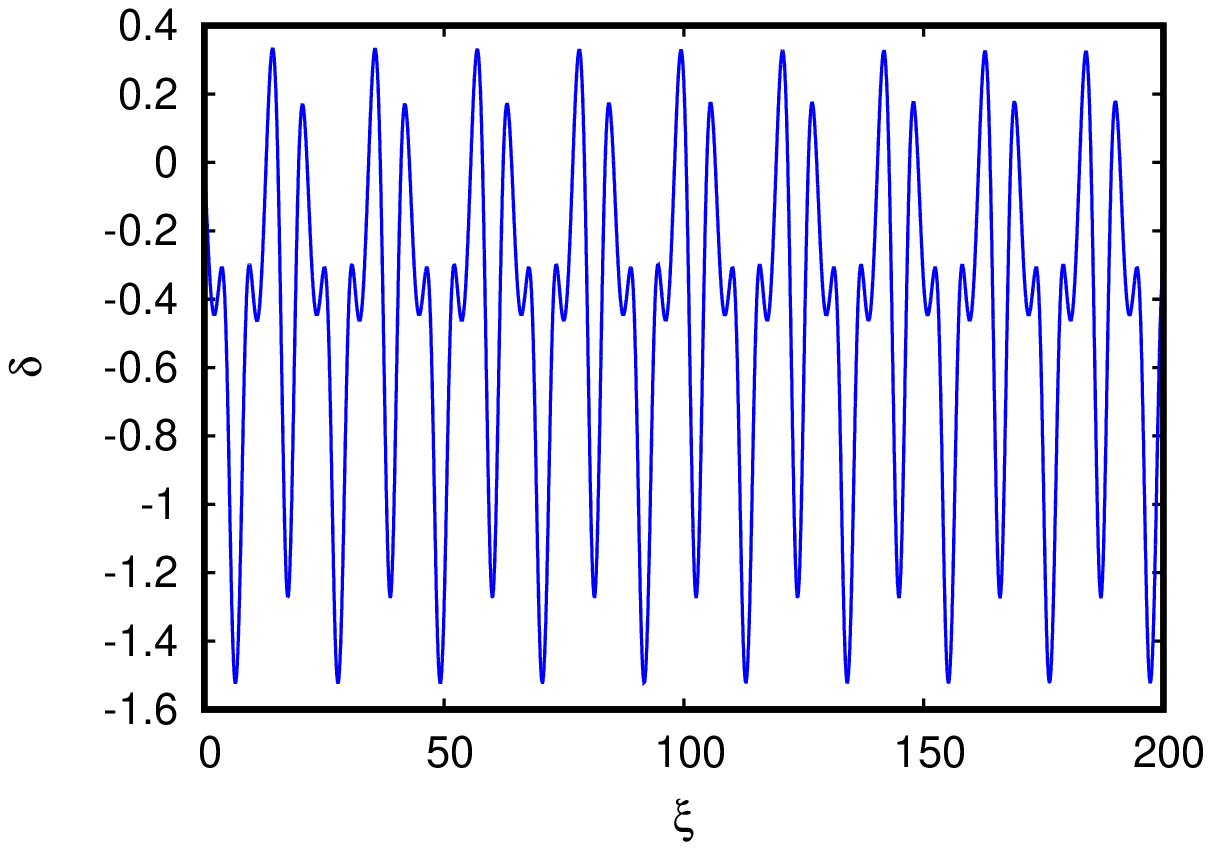}
\end{minipage}%
\begin{minipage}[c]{0.5\textwidth}
\includegraphics[width=3.in,height=2.in]{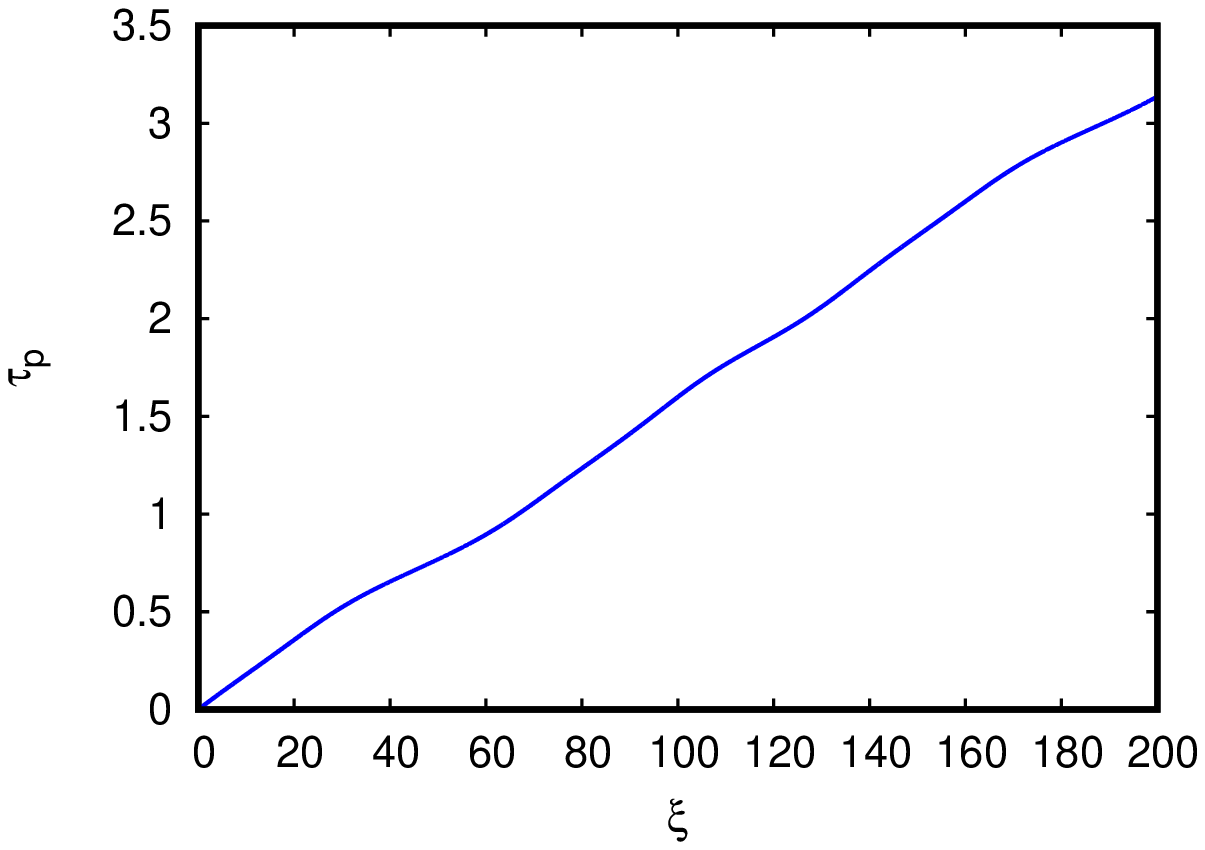}
\end{minipage}\\
\begin{minipage}[c]{0.5\textwidth}
\includegraphics[width=3.in,height=2.in]{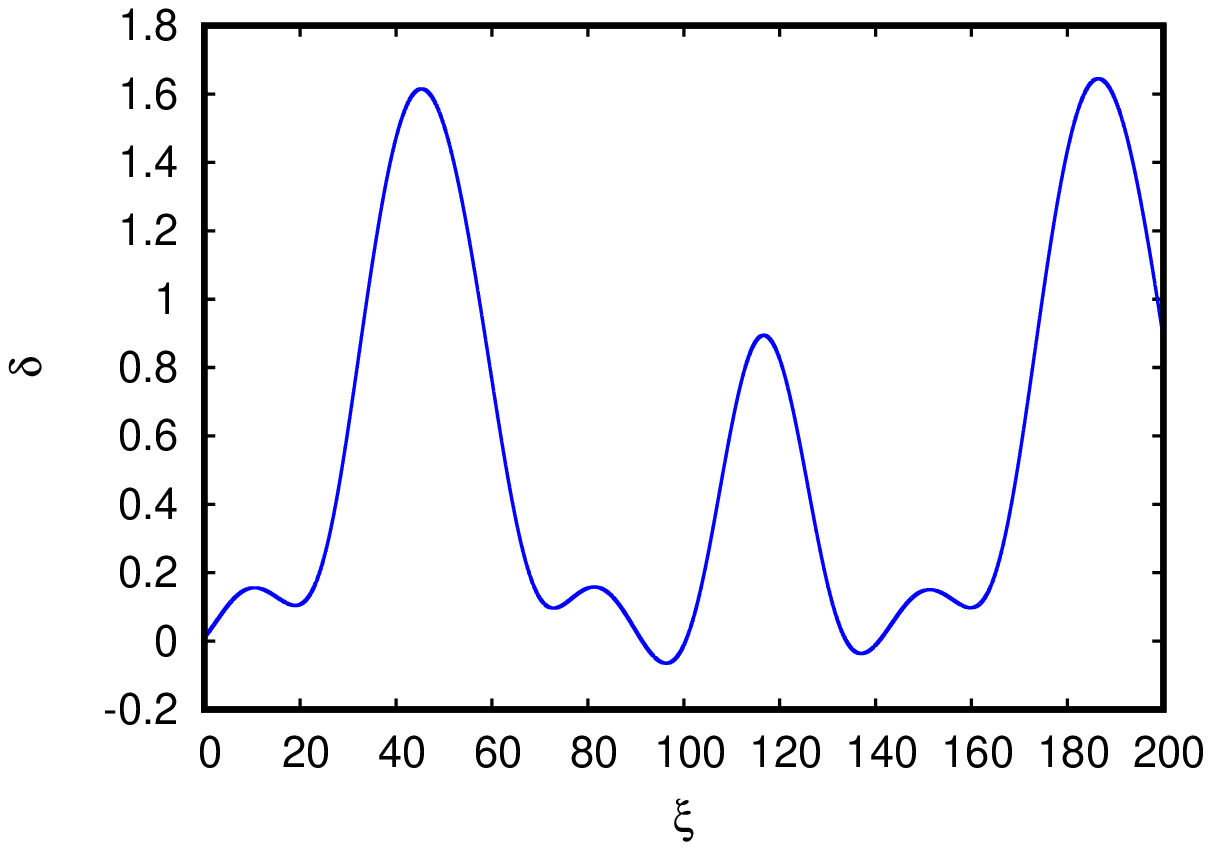}
\end{minipage}%
\begin{minipage}[c]{0.5\textwidth}
\includegraphics[width=3.in,height=2.in]{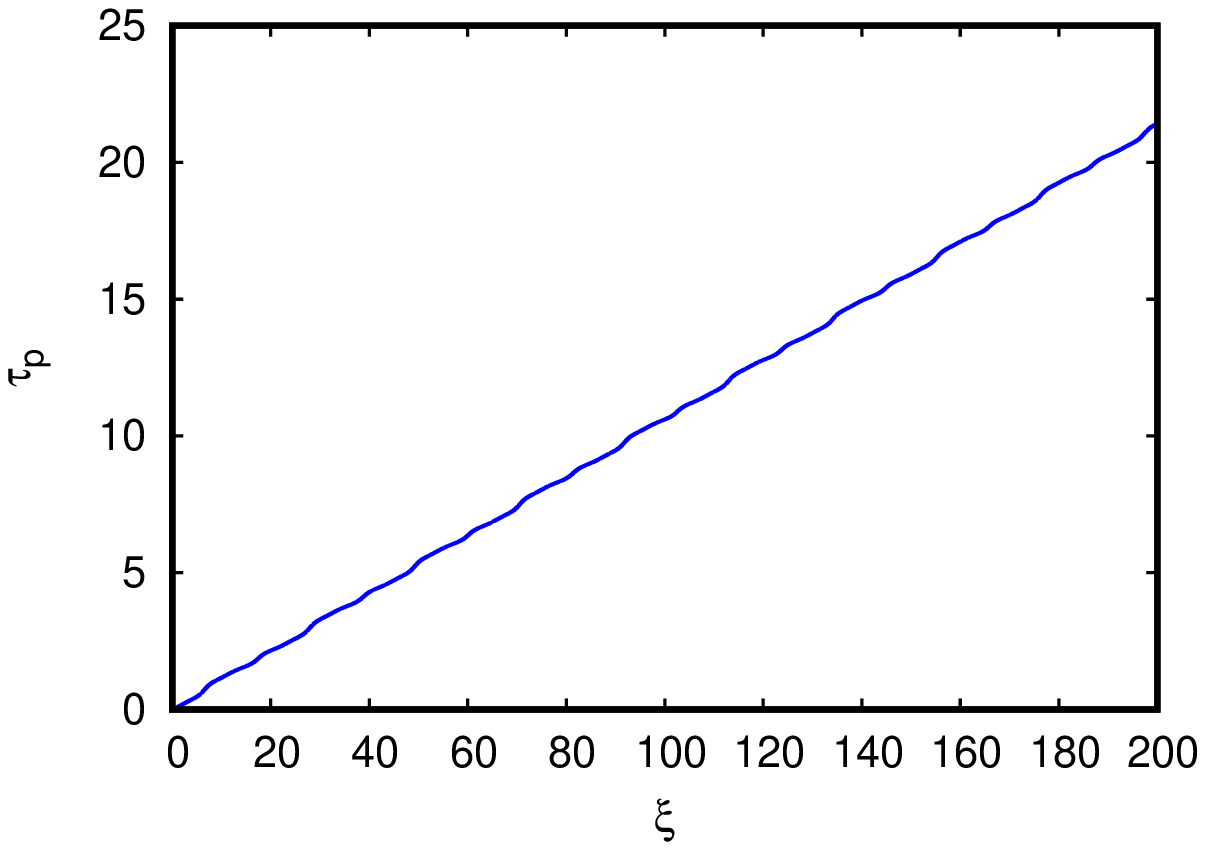}
\end{minipage}
\caption{\label{fig:three} Plots of $\delta$ (right graphs) and $\tau_p$ (left graphs) versus $\xi$, for $\kappa=0.97$, $\eta=0.75$ and $\tau_{sh}=0.1$. Top graphs: $dn$ mode, middle graphs: $cn$ mode, bottom graphs: $sn$ mode.}
\end{figure*}

\begin{figure*}\centering
\begin{minipage}[c]{0.5\textwidth}
\includegraphics[width=3.in,height=2.in]{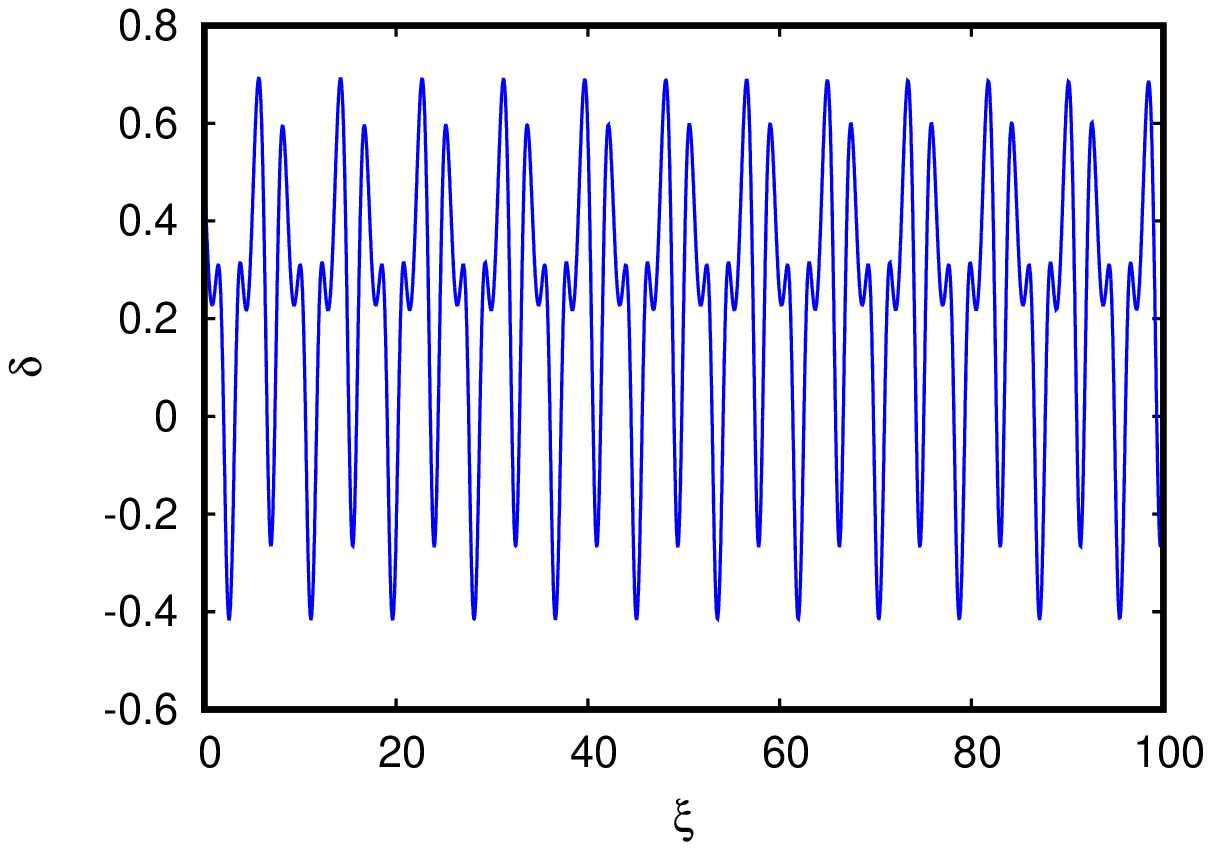}
\end{minipage}%
\begin{minipage}[c]{0.5\textwidth}
\includegraphics[width=3.in,height=2.in]{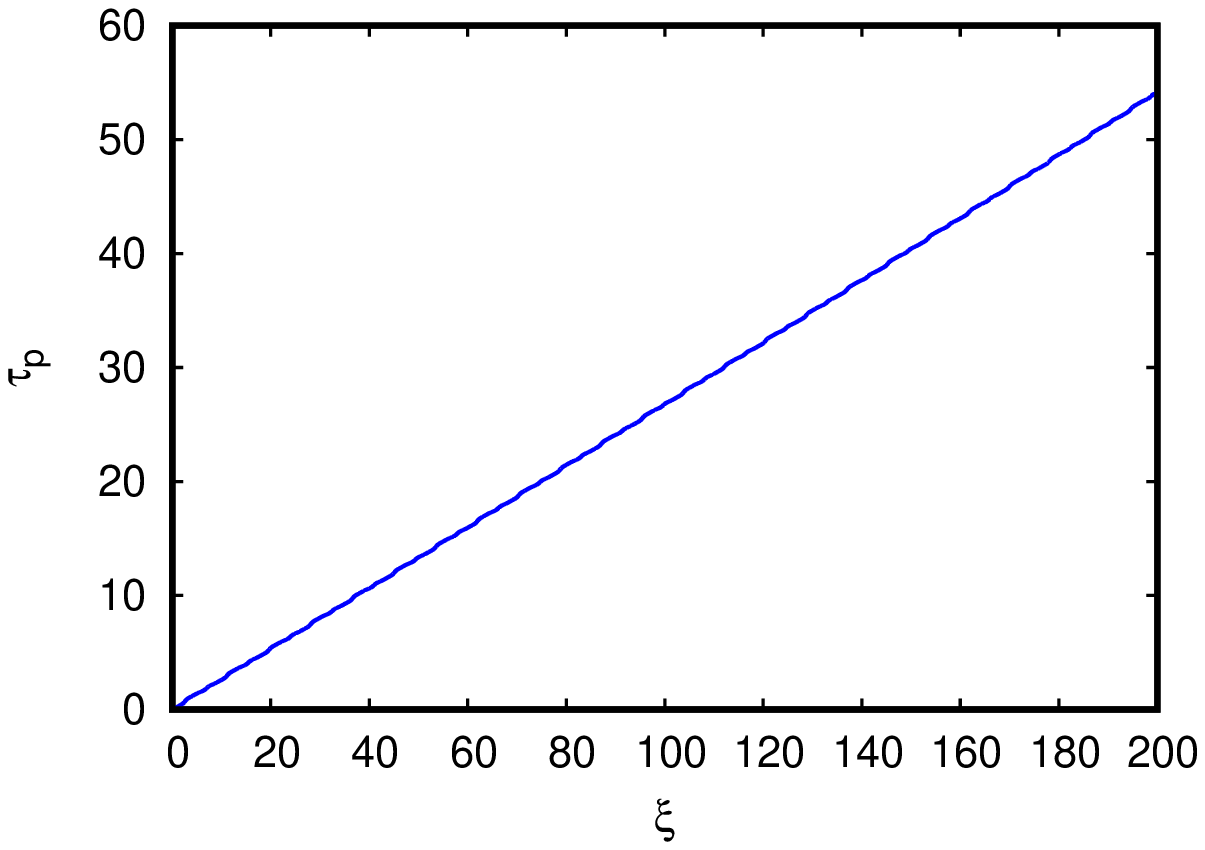}
\end{minipage}\\
\begin{minipage}[c]{0.5\textwidth}
\includegraphics[width=3.in,height=2.in]{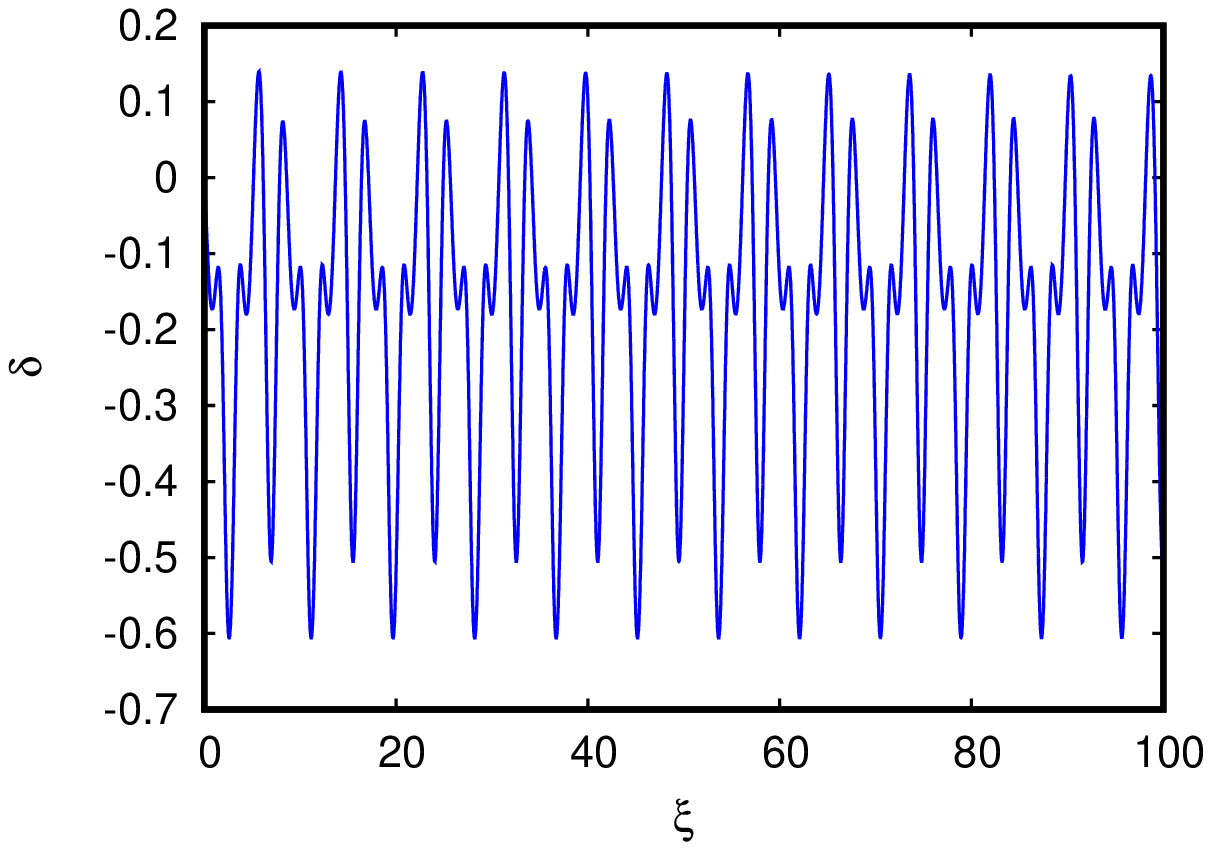}
\end{minipage}%
\begin{minipage}[c]{0.5\textwidth}
\includegraphics[width=3.in,height=2.in]{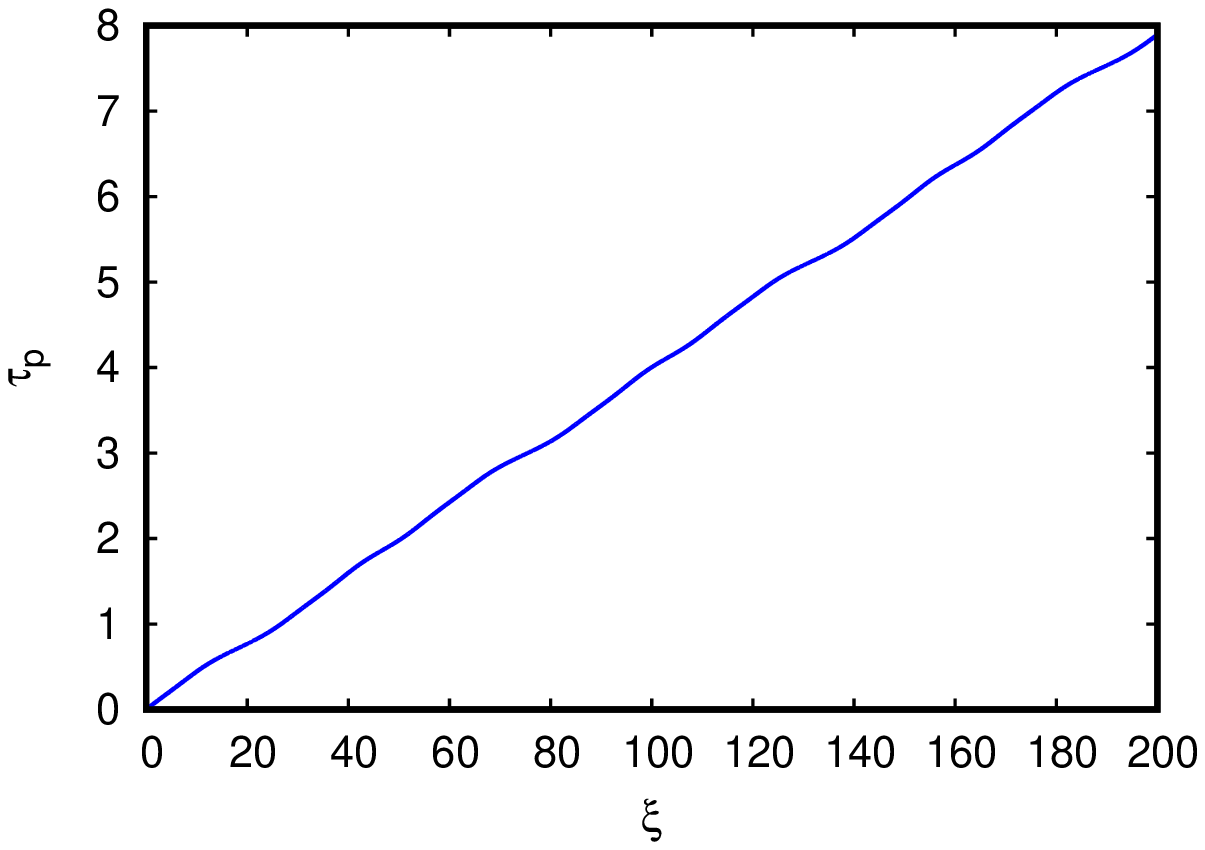}
\end{minipage}\\
\begin{minipage}[c]{0.5\textwidth}
\includegraphics[width=3.in,height=2.in]{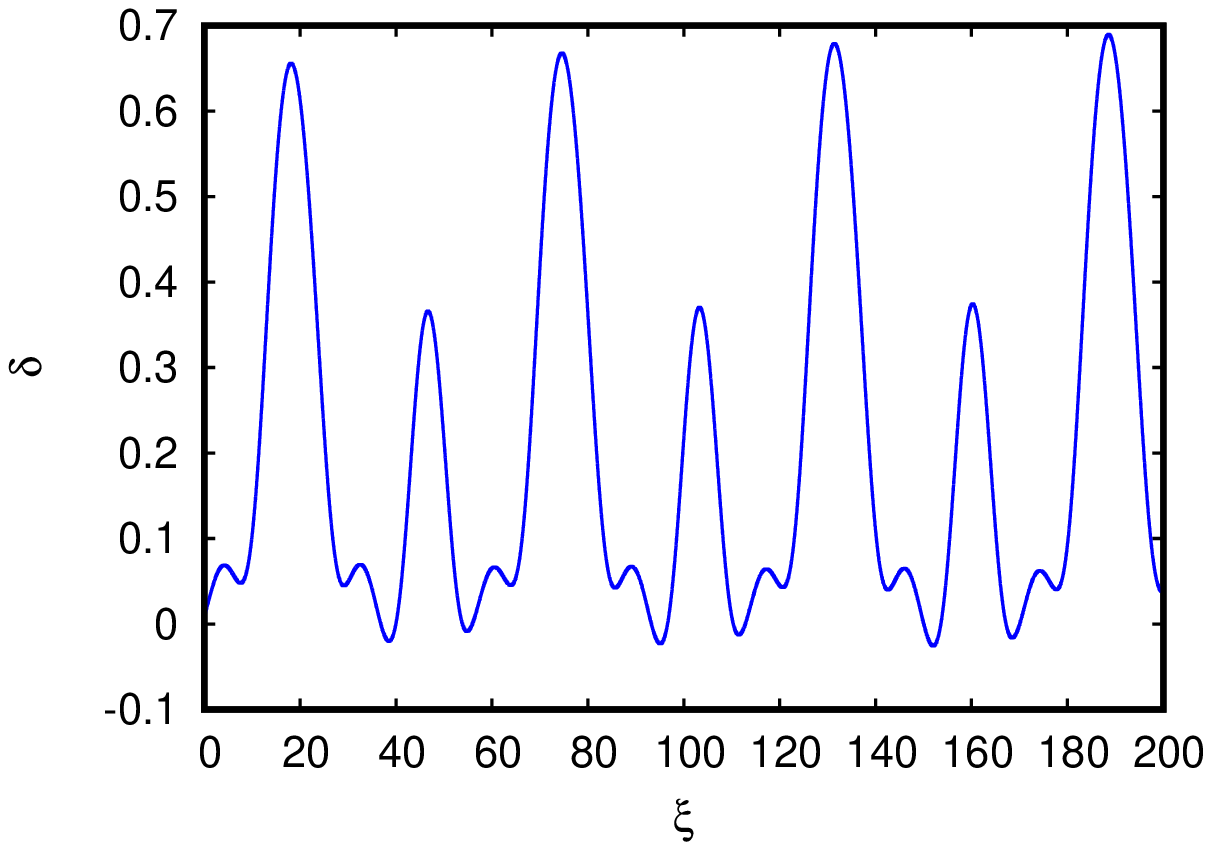}
\end{minipage}%
\begin{minipage}[c]{0.5\textwidth}
\includegraphics[width=3.in,height=2.in]{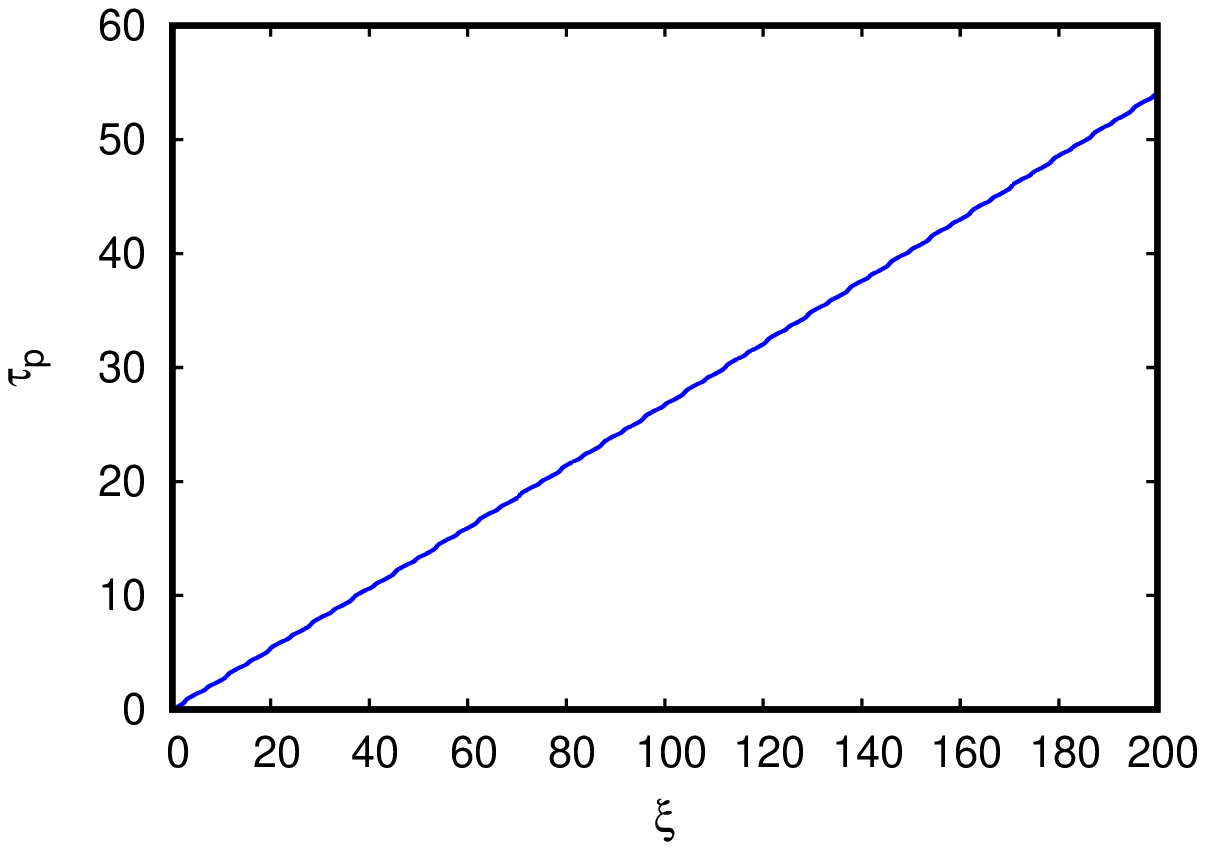}
\end{minipage}
\caption{\label{fig:four} Plots of $\delta$ (right graphs) and $\tau_p$ (left graphs) versus $\xi$, for $\kappa=0.97$, $\eta=0.75$ and $\tau_{sh}=0.25$. Top graphs: $dn$ mode, middle graphs: $cn$ mode, bottom graphs: $sn$ mode.}
\end{figure*}

Combining eqs.~(\ref{e3a}, (\ref{e3b}), (\ref{e4}) and (\ref{e5}) to obtain the Lagrangian \cite{t31}, and next applying the Euler-Lagrange formalism with respect to the two variational parameters, we obtain the evolution equations for the three boundstates as follows: 

\begin{enumerate}
 \item For $U_1(\tau, \xi) =\sigma dn\left[\sigma(\tau - \nu \xi) \right]
e^{i \frac{\eta^2}{2}\xi}$  and $U_2(\tau, \xi) = U_0 dn\left[\sigma(\tau - \tau_p) \right]
e^{-i \delta(\tau - \tau_p)}$:

\begin{eqnarray}
\frac{\partial \tau_p}{\partial \xi}&=& \frac{4 \tau_{sh} \sigma^3}{{\bf E}} \int^{{\bf K}}_{
0}{dn^2[\sigma(\tau - \tau_p)] dn^2\left[\sigma(\tau - \nu \xi)\right] d\tau}, \label{e6a} \\
\frac{\partial \delta}{\partial \xi}&=& -\frac{4(1 + 2 \tau_{sh} \delta) \kappa^2 \sigma^4}{{\bf E}}
\int^{{\bf K}}_{0}{cn[\sigma(\tau - \tau_p)]dn[\sigma(\tau - \tau_p)]sn[\sigma(\tau - \tau_p)]
dn^2\left[\sigma(\tau - \nu \xi)\right] d\tau}. \nonumber \\ \label{e6b}
\end{eqnarray}

\item For $U_1(\tau, \xi) =\sigma dn\left[\sigma(\tau - \nu \xi) \right]
e^{i \frac{\eta^2}{2} \xi}$ and $U_2(\tau, \xi) = U_0cn\left[\sigma(\tau - \tau_p)\right] 
e^{-i \delta(\tau - \tau_p)}$:

\begin{eqnarray}
\frac{\partial \tau_p}{\partial \xi}&=& \frac{4\kappa^2 \tau_{sh} \sigma^3}{\lbrack{\bf E}+(1-k^2){\bf K}\rbrack}
\lbrack\int^{{\bf K}}_{0}{dn^2[\sigma(\tau - \tau_p)] dn^2\left[\sigma(\tau - \nu \xi)\right]  d\tau} - \frac{\sqrt{2 - \kappa^2}}{\eta}(1 - \kappa^2){\bf E} \rbrack, \nonumber \\
\label{e7a} \\
\frac{\partial \delta}{\partial \xi}&=& -\frac{4 \kappa^4 \sigma^4(1 + 2 \tau_{sh}\delta)}{\lbrack{\bf E}+(1-k^2){\bf K}\rbrack}
 \int^{{\bf K}}_{0}{cn[\sigma(\tau - \tau_p)] dn[\sigma(\tau - \tau_p)] sn[\sigma(\tau - \tau_p)]dn^2\left[\sigma(\tau - \nu\xi)\right] d\tau}. \nonumber \\
 \label{e7b}
\end{eqnarray}

\item For $U_1(\tau, \xi) =\sigma dn\left[\sigma(\tau - \nu \xi) \right]
e^{i \frac{\eta^2}{2} \xi}$ and $U_2(\tau, \xi) = U_0 sn \left[\sigma(\tau - \tau_p) \right] 
e^{-i \delta(\tau - \tau_p)}$:

\begin{eqnarray}
\frac{\partial \tau_p}{\partial \xi}&=& \frac{4 \tau_{sh} \sigma^3}{({\bf E} - {\bf K})}
\lbrack\int^{{\bf K}}_{0}{dn^2[\sigma(\tau - \tau_p)] dn^2\left[\sigma(\tau - \nu \xi)\right]  d\tau} - \frac{\sqrt{2 - \kappa^2}}{\eta}{\bf E} \rbrack, \label{e8a} \\
\frac{\partial \delta}{\partial \xi}&=& -\frac{4 \sigma^4(1 + 2 \tau_{sh}\delta)}{({\bf E} - {\bf K})} \int^{{\bf K}}_{0} {cn[\sigma(\tau - \tau_p)] dn [\sigma(\tau - \tau_p)] sn[\sigma(\tau - \tau_p)] dn^2\left[\sigma(\tau - \nu\xi)\right] d\tau}. \nonumber \\
\label{e8b}
\end{eqnarray}

\end{enumerate}
where ${\bf E}=E(K,k)$ is the Elliptic integral of the second kind. None of the above three sets of two coupled equations can be solved analytically, and for this reason we resort to numerical simulations to find the evolutions of the two variational parameters as function of $xi$. For the numerical analysis, we used the $3/8$ Simpson rule to tackle the finite integrations and a sixth-order Runge-Kutta algorithm to solve the coupled first-order ordinary differential equations. The problem involves several parameters which we cannot consider all together for this would require an extended discussion. Therefore we only present results for some interesting set of values of paramaters, namely $\kappa=0.97$, $\eta=0.75$ and three different values of the self-steepening coefficient $\tau_{sh}$ here playing the role of perturbation. \\
Figs. \ref{fig:two}, \ref{fig:three} and \ref{fig:four} are plots of the soliton-train position shifts $\delta$ (right graphs) and phase shifts $\tau_p$ (right graphs) versus $\xi$, for $\tau_{sh}=0.08$, $0.1$ and $0.25$ respectively. the top graphs in each figure are variational parameters for the $dn$ boundstates, the middle graphs are for $cn$ and the bottom graphs are for the $sn$ boundstates. 

As one sees, the main dominant features are anharmonic oscillations of the three boundstates and a "stairecase-like" growth of the phase shift. The trajectories of the $dn$ and $cn$ boundstates are amost similar for the three valaues of $\tau_{sh}$ selected, but there is a sizable difference in the periods of their phase shifts. As we increase the self-steepening coefficient the periods of anharmonic oscillations of the three boundstates decrease, which is also the case for their phase shifts. \\
It is worth to stress that the motions of the three boundstates, as revealed from simulations, are quite complex for values of $\kappa$ (the common modulus of the three Jacobi Elliptic functions), though mainly dominated by anahrmonic oscillations of their position shifts and staircase-like variations of their phase shifts with $xi$. However, when $\kappa$ becomes too small (such that the three boundstates decay into harmonic fields), $\delta$ reduces to simple harmonic oscillations while $\tau_p$ becomes a linear function of its argument $\xi$. At reasonably high $\kappa$, the three boundstates display chaotic motions when $\tau_{sh}$ is relatively high. This later observation from numerical results, however, is to be taken with caution since strickly speaking the variational treatment was restricted to small values of $\tau_{sh}$. 

\section{\label{sec:four} Conclusion}
We proposed a new optical trapping scheme involving a periodic train of high-intensity pulses co-propagating with small-amplitude radiations, in a hollow-core photonic crystal fiber filled with a noble gas. The two fields are assumed to propagate in the same group-velocity dispersion channel, but with opposite circular polarizations such that their cross-phase modulation interaction is optimized. This leads to the small-amplitude radiations being always trapped in the field of the high-intensity pulse train, thus avoiding beeing wiped out along the fiber by dispersion. By constructing an artificial soliton-crystal structure from simple consideration of a temporal entanglement of a periodic continuous input train pulses, and exploiting the fact that such artificial structure also obeys the nonlinear Schr\"odinger equation, we considered its interaction with linear waves in terms of two wave equations coupled via a cross-phase modulation term. We found that the temporal profile of the linear-wave envelope was governed by a first-order Lam\'e equation, whose boundstate spectrum consists of exactly three modes which we determined unambiguously. One of the three modes was a "duplicate" of its pump, two were also-degenerate while a third mode was a train of optical dark solitons. Considering the effect of self-steepening on the pump and probe field propagations, we obtained that the pump soliton train was linearly shifted along the temporal scale while motions of the three boundstates were more complex, involving anharmonic oscillations and a staircase-like evolution of their phase shifts along the photonic fiber. The present results are interesting and offer new prospects for wave-guided-wave processes, wave duplication and quantum cloning \cite{syl,bin,jo} putting into play time-multiplexed high-intensity pulses and small-amplitude optical fields propagating together in photonic fiber media.   

\acknowledgments
The authors thank the Alexander von Humboldt (AvH) foundation for logistic supports. We dedicate this work to Professor Francis Kofi Ampenyi Allotey, a mentor.


\begin{thebibliography}{}\label{sec:references}

\bibitem{t1} G. P. Agrawal, \emph{Nonlinear Fiber Optics} (4th ed., Academic Press, San Diego, CA, 2007).
\bibitem{t2} L. F.Mollenauer, R. H. Stolen, and J. P. Gordon, Phys. Rev. Lett. {\bf 45}, 1095(1980).
\bibitem{t3}  A. M. Weiner, J. P. Heritage, R. J. Hawkins, R. N. Thurston, E. M. Kirschner, D. E. Leaird and W. J. Tomlinson, Phys. Rev. Lett. {\bf 61}, 2445(1988).
\bibitem{t4} L. F. Mollenauer, R. H. Stolen, J. P.Gordon and W. J. Tomlinson, Opt. Lett. {\bf 8}, 289(1983).
\bibitem{t4a} R. W. Boyd, \emph{Nonlinear Optics} (3rd ed., Academic Press, 2008).
\bibitem{m1} J. T. Manassah, Opt. Lett. {\bf 15}, 670(1990).
\bibitem{m2} J. T. Manassah, Opt. Lett. {\bf 16}, 587(1991).
\bibitem{m3} R. de la Fuente and A. Barthelemy, IEEE J. Quantum Electron. {\bf 28}, 547(1992).
\bibitem{t5} G. Genty, M. Lehtonen, and H. Ludvigsen, Opt. Express {\bf 12}, 4614(2004).
\bibitem{t6}  S. Trillo, S. Wabnitz, E. M. Wright, and G. I. Stegeman, Opt. Lett. {\bf 13}, 871(1988).
\bibitem{t7} Y. S. Kivshar, Opt. Lett. {\bf 17}, 1322(1992).
\bibitem{t8} A. V. Gorbach, D. V. Skryabin, J. M. Stone, and J. C. Knight, Opt. Express {\bf 14}, 9854(2006).
\bibitem{t9}  A. V. Gorbach and D. V. Skryabin, Phys. Rev. A {\bf 76}, 053803(2007).
\bibitem{t10} A. Hasegawa, Opt. Lett. {\bf 5}, 416(1980).
\bibitem{t13} P. St. J. Russell, Science {\bf 299}, 358(2003).
\bibitem{t14} J. C. Travers, W. Chang, J. Nold, N. Y. Joly and P. St. J. Russell, J. Opt. Soc. Am. B{\bf 28}, A11(2011).
\bibitem{t15} D. S. Mbieda Petmegni and A. M. Dikand\'{e}, J. Mod. Opt. {\bf 64}, 1192(2017).
\bibitem{steig1} D. Rand, K. Steiglitz and P. R. Prucnal, Phys. Rev. A{\bf 72}, 041805(R)(2005).
\bibitem{steig2} C. Ottaviani, S. Rebic, D. Vitali, and P. Tombesi, Phys. Rev. A{\bf 73}, 010301(R)(2006).
\bibitem{steig3} K. Steiglitz and D. Rand, Phys. Rev. A{\bf 79}, 021802(R)(2009). 
\bibitem{t12} A. M. Dikand\'e, Phys. Rev. A {\bf 81}, 013821(2010).
\bibitem{t16} N. Y. Joly, J. Nold, W. Chang, P. H\"olzer, A. Nazarkin, G. K. L. Wong, F. Biancalana and P. St.J. Russell, Phys. Rev. Lett. {\bf 106}, 203901(2011).
\bibitem{t17} J. C . Knight, Nature {\bf 424}, 847(2003).
\bibitem{t18} P. St. J. Russell, J. Lightwave Technol. {\bf 24}, 4729(2006).
\bibitem{moha} M. F. Saleh and F. Biancalana, Phys. Rev. A{\bf 84}, 063838(2011).
\bibitem{fac} M. Fac\~ao, M. I. Carvalho and P. Almeida, Phys. Rev. A{\bf 87}, 063803(2013).
\bibitem{moha1} M. F. Saleh and F. Biancalana, J. Opt. {\bf 18}, 013002(2016).
\bibitem{t19} D. S. Mbieda Petmegni, A. M. Dikand\'e and B. Z. Essimbi, Appl. Phys. B{\bf 123}, 171(2017).
\bibitem{t20} W. Chang, A. Nazarkin, J. C. Travers, J. Nold, P. H\"olzer, N. Y. Joly and P. St. J. Russell, Opt. Express {\bf 19}, 21018(2011).
\bibitem{t21} M. F. Saleh and Fabio Biancalana, Phys. Rev. A{\bf 87}, 043807(2013).
\bibitem{t22} M. F. Saleh, W. Chang, J. C. Travers, P. St.J. Russell, and F. Biancalana, Phys. Rev. Lett. {\bf 109}, 113902(2012).
\bibitem{mar} F. DeMartini, C. H. Townes, T. K. Gustafson, and P. L. Kelley, Phys. Rev. {\bf 164}, 312 (1967).
\bibitem{stepe1} M. Trippenbach and Y. B. Band, Phys. Rev. A{\bf 57}, 4791(1998)
\bibitem{stepe2} J. Moses, B. A. Malomed and F. W. Wise, Phys. Rev. A{\bf 76}, 021802(2007).
\bibitem{stepe3} J. Fujioka and A. Espinosa, Chaos {\bf 25}, 113114(2015).
\bibitem{luth} H. Luther, Math. Comp. {\bf 22}, 434(1968). 
\bibitem{men1} P. K. A. Wai, C. R. Menyuk, and B. Raghavan, J. Lightwave Technol. {\bf 14}, 1449(1996).
\bibitem{malomed} B. A. Malomed, Phys. Rev. B{\bf 38}, 9242(1988).
\bibitem{t23} D. Jr. Fandio Jubgang, A. M. Dikand\'e and A. Sunda-Meya, Phys. Rev. A{\bf 92}, 053850(2015).
\bibitem{t31} D. Jr. Fandio Jubgang and A. M. Dikand\'e, J. Opt.Soc. Am. B{\bf 34}, 2721(2017).
\bibitem{t26} M. Abramowitz and I. A. Stegun, \emph{Hand Book of Mathematical Functions with Formulas, Graphs and Mathematical Tables} (10th ed., National Bureau of Standards, 1972).
\bibitem{t27} F. Bowman, \emph{Introduction to Elliptic Functions with Applications} (Dover, 1961).
\bibitem{t28} F. M. Arscott and I. M. Khabaza, \emph{Tables of Lam\'e polynomials, mathematical tables series} (Pergamon Press, Oxford, 1962).
\bibitem{t29} A. M. Dikand\'e, Phys. Scrip. {\bf 60}, 293 (1999).
\bibitem{bej} P. B\'ejot, B. E. Schmidt, J. Kasparian, J. P. Wolf and F. Legar\'e, Phys. Rev. A{\bf 81}, 063828(2010).
\bibitem{chad} C. Husko and P. Colman, Phys. Rev. A{\bf 92}, 013816(2015).
\bibitem{rai} F. Raineri, T. J. Karle, V. Roppo, P. Monnier and R. Raj, Phys. Rev. A{\bf 87}, 041802(R)(2013).
\bibitem{syl}S. Fasel, N. Gisin, G. Ribordy, V. Scarani and H. Zbinden, Phys. Rev. Lett. {\bf89}, 107901(2002).
\bibitem{bin}B. Qi and L. Qian, Opt. Lett. {\bf 32}, 418(2007).
\bibitem{jo}J. Burkart, T. Sala, S. Kassi, D. Romanini and M. Marangoni, Opt. Lett. {\bf 40}, 816(2015).

\end{thebibliography}
\end{document}